\documentclass[12pt]{article}
\usepackage{amssymb}
\usepackage{amsmath}
\usepackage{mathrsfs}
\usepackage{multirow}
\newcommand{\nn}{\nonumber}
\newcommand{\qed}{\hfill \rule{1.3mm}{3mm}}
\begin{document}

\title{\bf  Understanding Cross-sectional Dependence in Panel Data} \vspace{1cm}

\author{Gopal K Basak\thanks{ Stat-Math Unit,  Indian Statistical Institute,  Kolkata, India,  
email: gkb@isical.ac.in} \;and   Samarjit Das\thanks{Corresponding author:  ERU, Indian Statistical Institute, 
203 B.T. Road, Kolkata-700108, India, Tel: +91-33-2575-2627, email: samarjit@isical.ac.in}\\ Indian Statistical 
Institute \\
}

\date{}
\maketitle
\begin{abstract}   We  provide various norm-based definitions of different types of cross-sectional dependence 
and the relations between them. These definitions facilitate  to comprehend and to characterize the various forms 
of cross-sectional dependence, such as strong, semi-strong, and weak dependence. 
Then we  examine the asymptotic properties of  parameter estimators  both 
for  fixed (within) effect estimator and random effect (pooled) estimator for  linear panel data models  
incorporating various forms of cross-sectional dependence. The asymptotic properties are also derived when both 
cross-sectional and temporal dependence are present. Subsequently,  we develop consistent and  robust  standard 
error  of the parameter estimators both for fixed effect and random effect 
model separately. Robust standard errors are developed (i) for pure cross-sectional dependence; and (ii) also for 
cross-sectional and time series dependence.  
Under  strong or semi-strong cross-sectional dependence, it is established that when the  time dependence comes 
through the idiosyncratic errors, such time dependence  does not have any influence in the asymptotic variance 
of $(\hat{\beta}_{FE/RE}). $  Hence, it is argued that in estimating $Var(\hat{\beta}_{FE/RE}),$  
Newey-West kind of correction injects bias in the variance estimate.   
Furthermore, this article lay down  conditions under which $t$, $F$  and the $Wald$ statistics based on the 
robust covariance matrix estimator give valid inference.

\noindent 
\end{abstract}

\vspace{0.2cm}

\vspace{0.2cm}

\newpage

\baselineskip0.6cm

\section{Introduction}

Over the past few years there has been a growing literature on the problem of error cross section dependence in panel regressions.    Such dependence may come due to a number of counts ; viz.,  
  selecting individuals non-randomly, unobserved common shocks, due to a single currency, due to common aggro-climatic environment; and
policies adopted by the central authority and so on. Stephan (1934) argues that  ``in dealing with social
data, we know that by virtue of their very social character, persons, groups and their
characteristics are interrelated and not independent."

In the presence of spatial  dependence, Baltagi and Pirotte (2010) show that the
test of hypothesis based on the standard panel data estimators that ignore spatial dependence can
lead to misleading inference.  Correlation across units in panels has also serious drawbacks (O'Connell, 1998)  on commonly used panel
unit root tests, since several of the existing tests assume independence (Levin, Lin and Chu, 2002; Im,
Pesaran and Shin, 2003).  Such dependence has serious implication in testing for convergence  hypothesis (Banerjee { \it et al.}, 2005). Basak and Das (2017) showed that Chow type F-test are severely oversized when such dependence is present in panel data. Moreover, they also evidenced that Hausman test is quite unstable across various type of cross-sectional dependence.

In this paper, we first attempt to define various forms of dependence across units  in a comprehensive way. Previously, several researchers have attempted to characterize cross-sectional dependence   based on either cross-sectional or both cross-sectional and time  average series ; and   may be  in a scatter way. Forni and Lippi (2001) define two different type of cross-sectional dependence, viz., idiosyncratic (weak dependence) and common factor (strong dependence). Anderson et al. (2009) propose definitions of weak and strong cross-sectional dependence  based on eigenvalues of spectral density. The purpose of these work was to propose method for dimension reduction; not to study large sample properties of slope parameters in panel data models. Chudik et al (2011) defines various forms 
of cross-sectional dependence based on the cross-sectional  weighted average  where weights satisfy some 
technical conditions called `granularity condition'.  This granularity condition facilitates greatly to 
derive the asymptotic properties; but its physical interpretation remains unclear. 
This kind of transformation  certainly 
advantageous  for  the derivation of the asymptotic properties though it looses the cross-sectional information. 
Moreover, the kind of assumptions made on the  cross-sectionally averaged single time series put restrictions on 
the form of cross-sectional dependence (Vogelsang, 2012), are generally not met in  many practical situations.
 Further, the choice of weights 
corresponding to granularity condition is somewhat arbitrary, and subsequently the quality of parameter estimators may substantially depend on the choice of the weights. Under this backdrop, we feel that there is a need to define various forms of cross-sectional dependence across units in a comprehensive way.
It is shown that our definition of cross-sectional dependence includes spatial dependence (Anselin, 2001); error 
factor structure of Bai (2009) and Pesaran (2006);   Chudik et al (2011)'s strong factor, semi-strong factor, weak 
factors; and also other kind of dependence.   We provide several real life examples which help  to 
comprehend various forms of dependence.

We then attempt to find answer to a few practical and relevant questions. Are the fixed effect and random effect estimators consistent under various forms of cross-sectional and time series dependence? Can we find CLT for such estimators? Is it possible to find a robust standard error; robust for both cross-sectional and time series dependence? Finding answer of these questions are very important as applied researchers are using these estimators on regular basis. Here it may 
be mentioned that  some of these asymptotic results are, at best, partly 
known. The  area is quite wide open; the literature does not have a satisfactory theoretical foundation.   The work of Driscoll and  Kraay (1998) is the first attempt to find a proper robust standard error. They considered random effect model and a particular kind of dependence which may be termed as strong dependence. Vogelsang (2012) study the properties of fixed effect
model under finite $N$  or under the case which may be termed as weak dependence.  Gon\c{c}alves (2011) proposed moving block bootstrap method to find robust standard error assuming CLT. Our asymptotic results for fixed and random effect slope 
estimators cover separately  for (i)  cross-sectional dependence of all forms; 
and (ii)  cross-sectional and time series dependence.  Separate study for (i) and (ii) gives more insights on the asymptotic behaviour of the estimators. For example, we argue in this paper that Newey-West kind of correction  as 
used in Driscoll and  Kraay (1998) is irrelevant in many practical situations  when cross-sectional dependence is strong or semi-strong.
Hence, in  estimating $Var(\hat{\beta}_{RE}),$  Newey-West kind of correction as considered 
in Driscoll and  Kraay (1998) injects bias in the variance estimate.

The remainder of the paper is organized as follows. The model and various kinds of cross-sectional dependence in 
error are provided in Section 2. Several examples are given to comprehend the broad scope of such definitions. 
Results are given to show how different forms of dependence are interconnected.   Section 3 has two subsections. 
Here asymptotic results are presented (i) for pure cross-sectional dependence; and (ii) also for cross-sectional 
and time series dependence. Subsection 3.1 provides asymptotic results for fixed and random effect slope 
estimators under the presence of pure cross-sectional dependence. 
Subsection 3.2 obtains  asymptotic results for fixed and random effect slope estimators under both for 
cross-sectional and time series dependence.  
The paper concludes in Section 4. An appendix provides the proofs of the asymptotic results.

\section{Defining Cross-sectional Dependence}

Let $y_{it}$ be the observation on the $i$-th cross-section unit at time $t$  and $x_{it}$ be a $k \times 1$ vector of observed
individual-specific regressors on the $i$-th cross-section unit at time $t$ for $i =
1, 2, ..., N, t = 1, 2,..., T$.  Consider a very general panel data model as 
\begin{equation*}
y_{it}=f( x_{i,t}, \theta) + \varepsilon _{it}.   
\end{equation*}
This regression is as general as possible and it considers  non-stationary variables as well. 
The error vector
$\varepsilon_{.t}= [\varepsilon_{1t},\ldots,\varepsilon_{Nt}]'$ is assumed to be 
i.i.d.  with $E(\varepsilon_{.t})=0$ and
$E(\varepsilon_{.t}\varepsilon_{.t}')=\Omega$.  Let $\lambda_1, \lambda_1, \ldots, \lambda_N $ are $N$ eigenvalues, in ascending order, of $\Omega.$

\vspace{0.2cm}
\noindent{\bf Weak Cross-sectional Dependence:} Dependence across individuals is said to be weak when $\lambda_N  =  O_p(1).$

Therefore, weak dependence implies all eigenvalues of $\Omega$ are finite. Independence  is regarded as weak dependence
\vspace{0.2cm}

\noindent{\it Example 1:} Any diagonal $\Omega$ matrix  necessarily implies weak dependence.
\vspace{0.2cm}

\noindent{\it Example 2:} Any band $\Omega$ matrix  of finite width  necessarily implies weak dependence.
\vspace{0.2cm}

\noindent{\it Example 3:} Any block diagonal $\Omega$ matrix with bounded (fixed-finite) block size  necessarily implies weak dependence. One possible way of dependence arises when individuals are correlated  within a cluster  but uncorrelated (or independent) outside the cluster. Here we are assuming cluster size does not increase but number of cluster increases with $N$.
\vspace{0.2cm}

\noindent{\it Example 4:} Weak dependence can also hold for dependence that decays sufficiently fast as observations become more distant according to some measure. Consider  a correlation matrix ($R$) corresponding to $\Omega.$  Note that $\lambda_{max} \Omega \le \max_{i} \omega_{ii} \lambda_{max} R.$  As for example, if $ r_{i,j}$ decays faster than $ \frac{1}{ |i-j| log|i-j|^p}, p >1  $ for large $|i-j|,$ dependence will be weak. 
\vspace{0.2cm}

\noindent{\it Example 5:} Consider any spatial weight matrix with both row and column summability.   Spatial dependence in AR, MA, ARMA structure with  such weight matrices   is an example of weak dependence.

\vspace{0.2cm}

\noindent{\bf Intermediate or Moderate Cross-sectional Dependence:} Dependence across individuals is said to be moderate when following condition holds:
$\lambda_{N}=O(h_N) , $ where $h_N \uparrow \infty$ as $N\uparrow \infty, $ but $\frac{h_N}{N} \to 0$ as $N \to \infty.$  

\vspace{0.2cm}
\noindent{\it Example 6:} As compared to Example 4  above, here $r_{i,j}$ must go to zero as $ |i-j| \to \infty$ at a slower rate; e.g., $p \le 1.$
\vspace{0.2cm} 

\noindent{\it Example 7:} Any band $\Omega$ matrix of width $h_N$ 
may give intermediate  dependence (e.g., with a positive constant diagonal (say 1) 
and positive contant off-diagonal element within the band (say $b (<1)$).

\vspace{0.2cm}
\noindent{\it Example 8:} Any block diagonal $\Omega$ matrix with largest block size $h_N$ where dependence
within the largest block may be maximum (e.g., contant diagonal entry, say 1, and a positive constant off-diagonal 
entry, say $b (<1)$) implies intermediate 
dependence. Here we are assuming that cluster size increases with $N$, so is the number of clusters.

\vspace{0.2cm}

\noindent{\bf Strong Cross-sectional Dependence:} Dependence across individuals is said to be strong when following
condition holds: $\lambda_{N}=O(N).$

\vspace{0.2cm}
\noindent{\it Example 9:} As compared to Example 4  above, here $r_{i,j}$ does not go to zero as 
$ |i-j| \to \infty. $ 

\vspace{0.2cm}

\noindent{\it Example 10:} Any block diagonal $\Omega$ matrix with fixed (say $m$ ) number of blocks
may give strong dependence (e.g., elements within the blocks are as in Example 8).  
Here we are assuming that number of cluster is fixed, but the each cluster size increases with $N$.
 
\vspace{0.2cm}

\noindent{\it Example 11:} Consider the factor representation ( see for example, Pesaran, 2006; Bai, 2009) of 
$\varepsilon_{.t}$ as  $\varepsilon_{.t}= \Lambda f_t+ u_{.t} $, $\Lambda$ is an $N \times m$ matrix of factor 
loadings.  The idiosyncratic error $ u_{.t}= [u_{1t},\ldots,u_{Nt}]'$ is an $N \times 1$ vector of i.i.d. errors  
with $E(u_{.t})=0$ and $E(u_{.t} u_{.t}')=\Sigma$; a finite with positive entity  diagonal matrix. It is easy to 
see that if  $ lim_{N \rightarrow \infty} (\frac{\Lambda' \Lambda} {N})$ is a positive definite matrix, then 
$\lambda_{N}=O(N),$  hence the dependence is strong. 

\vspace{0.2cm} 

\noindent{\it Note  for Example 11:}  $\Sigma$ does not need to be a diagonal matrix. The idiosyncratic error, 
$u_{.t}$ does not need to be independent across $t$, any stationary AR structure may be applicable.   

\vspace{0.5cm}

\noindent{\it Proposition 1:} {\it  The dependence across individuals is strong  iff  $\varepsilon_{.t}$
 contains factor structure as   $\varepsilon_{.t}= \Lambda f_t+ u_{.t} $, $\Lambda$ is an $N \times m$ matrix
 of factor loadings with $ lim_{N \rightarrow \infty} (\frac{\Lambda' \Lambda} {N})$ is a positive definite
 matrix. The  $ u_{.t}= [u_{1t},\ldots,u_{Nt}]'$ is an $N \times 1$ vector of i.i.d.  errors  with
 $E(u_{.t})=0$ and $E(u_{.t} u_{.t}')=\Sigma.$  }  
\vspace{0.2cm}

\noindent{\sc Proof:} Proof of  if part is  somewhat available  in the literature. However, for completeness 
purpose, a proof is given below.
Note that $ \Omega =Cov (\varepsilon_{.t})=\Lambda  \Lambda ^{\prime} +\Sigma. $    Since  
$  (\Lambda \Lambda^{\prime}) $ and $\Sigma$ both are non-negative definite matrices, 

$  \lambda_{max} (\Lambda \Lambda^{\prime}) \le  \lambda_{max} (\Lambda \Lambda^{\prime} +\Sigma) \le   \lambda_{max} (\Lambda \Lambda^{\prime}) + \lambda_{max} (\Sigma). $ 

Note that  $\lambda_{max} (\Sigma)$ is $O(1), $ and  
$\lambda_{max} (\Lambda \Lambda^{\prime})=\lambda_{max} (\Lambda^{\prime} \Lambda)=O(N)= \lambda_{min} (\Lambda \Lambda^{\prime}),$ 
since  $ lim_{N \rightarrow \infty} (\frac{\Lambda' \Lambda} {N})$ is a positive definite matrix. 
Thus $\lambda_{max}(\Omega)=\lambda_{max} (\Lambda \Lambda ^{\prime} +\Sigma)=O(N).$

Proof of only if part: Suppose there are $m$ eigenvalues, $\lambda_{N-i}, i=0, 1,\ldots, m-1 $ which are 
$O(N)$ and others are $o(N)$. Consider the spectral  decomposition of $\Omega$ as 

\begin{eqnarray}
\Omega &=& \sum_{i=1} ^{N} \lambda_i   P_i P_i ^{\prime}= \sum_{i=1} ^{N-m} \lambda_i   P_i P_i ^{\prime} + \sum_{i=N-m+1} ^{N} \lambda_i   P_i P_i ^{\prime} \nn\\
&=& \sum_{i=1} ^{N-m} \lambda_i   P_i P_i ^{\prime} + \sum_{i=N-m+1} ^{N} (\lambda_i  -c_i \lambda_1)  P_i P_i ^{\prime} + \sum_{i=N-m+1} ^{N} (c_i \lambda_1)  P_i P_i ^{\prime},  \;\; 0 <c_i \le 1,   \nn\\
&=& \{ \sum_{i=1} ^{N-m} \lambda_i   P_i P_i ^{\prime}  +  \sum_{i=N-m+1} ^{N} (c_i \lambda_1)  P_i P_i ^{\prime} \} +  \sum_{i=N-m+1} ^{N} (\lambda_i  -c_i \lambda_1)  P_i P_i ^{\prime} . \nn\\
&=& \sum_{i=1} ^{N} \lambda_i^{*}   P_i P_i ^{\prime}  + \sum_{i=1} ^{m} \delta_i V_i V_i ^{\prime} ; \;\;  \nn
\end{eqnarray}
where, $\lambda_i^{*} =\lambda_i , \; i=1, \ldots ; N-m, \;\;  \lambda_i^{*} = c_i\lambda_1 \; i=N-m+1, \ldots ; N,$
and,  $\;\; \delta_i=\lambda_{N-m+i}  - c_{N-m+i} \lambda_1 $ and  $ V_i=P_{N-m+i}, \; i=1, \ldots m. $
Now, $ \sum_{i=1} ^{N} \lambda_i^{*}   P_i P_i ^{\prime} =\Sigma$, and  $ \sum_{i=1} ^{m} \delta_i V_i V_i ^{\prime} = \Lambda \Lambda^{\prime}, \; \Lambda=[ V_1, V_2, \ldots V_m] \times D,   $  where 
$D$ is a diagonal matrix with $i$-th diagonal  is $\sqrt {\delta_i}.$
Note that  $\Sigma$ is  a positive definite matrix with all the eigenvalues are of $o(N), $ and $ \Lambda \Lambda^{\prime}$ is  a non-negative definite matrix with exactly $m$ number of  positive   eigenvalues with each of $O(N)$;  and $N-m$  zero eigenvalues. \qed

\vspace{0.5cm}

\noindent{\it Remark 1:}  The closer look of the above proof suggests that $\Sigma=P D^{*} P'$ has $N-m$ 
eigenvalues; some of which may be bounded and some of which  may be of $O(h_N),$ 
where$ D^{*}=Diag (\lambda_1,\ldots \lambda_N).$  Suppose there are $m_i, \; i=1\ldots, k $ eigenvalues which are of 
$O(h_{N}^i), i=1,2\ldots, k,$ and remaining eigenvalues are $O(1).$ For these $m_i$ eigenvalues we can have again 
factor structure with different kind of factor loading ($\Gamma_i$) such that 
$ lim_{N \rightarrow \infty} (\frac{\Gamma_i' \Gamma_i} {h_{N}^i}) $ is a finite positive definite matrix. Here 
$m_i ,$  for some or for all $i$, may tend to infinity  as $N\uparrow \infty. $ Chudik et al (2011) defines such 
factors as semi-weak or semi strong, although in a very restrictive sense. Similarly, corresponding to bounded 
eigenvalues, one can define factor structure with absolutely summeable factor loadings. Such factors are termed 
as weak factors (Chudik et al, 2011).

\subsection{Other Ways of Defining Cross-sectional Dependence}

In the previous section, we have defined various forms of dependence in terms of the maximum eigenvalue of $\Omega$. We have considered eigenvalues as these  are special implicit (continuous) functions of the elements of $\Omega=\{\omega_{i,j} \}$ and hence properties of $\{\omega_{ij}\}$ are embedded in eigenvalues. Moreover, eigenvalue theory is extremely developed. It may be interesting to define various forms of dependence in some other form of functions of $\{\omega_{ij}\}$. Some such functions may be $(i) \;\; \max_{j}(\sum_{i=1}^{N}|\omega_{ij}|), (ii) \;\; \sqrt(\frac{1} {N} \sum\sum \omega_{ij}^2) $ and $(iii) \;\; \frac{1} {N} \sum\sum |\omega_{ij}|. $ Similar to the above definitions of  dependence, we can define various dependence structure by using these three measures (related to norms). These popularly known norms (without the divisor $N$) are called {\it maximum absolute row sum norm,  Euclidean norm} and {\it taxicab norm}, respectively (Lewis, 1991). Some other norms,{\it viz}.,  the {\it Cartesian norm} may also be considered.

\vspace{0.2cm}
\noindent{\bf Weak Cross-sectional Dependence:} Dependence across individuals is said to be weak when either of these three following conditions holds:

$$(i) \;\; \limsup_{N}\max_{i}(\sum_{j=1}^{N}|\omega_{ij}|) < \infty , $$ 
$$(ii) \;\;  \;\; \limsup _{N} \sqrt{\frac{1} {N} \sum\sum \omega_{ij}^2} < \infty , $$ 
$$(iii) \;\;  \;\; \limsup _{N} \frac{1} {N} \sum\sum |\omega_{ij}|< \infty . $$ 

\vspace{0.2cm}

\noindent{\bf Intermediate or Moderate Cross-sectional Dependence:} Dependence across individuals is said to be moderate when either of these three following conditions holds:
$$(i) \;\; \max_{i}(\sum_{j=1}^{N}|\omega_{ij}|) = O(h_N) , $$ 
$$(ii) \;\;  \;\; \sqrt{\frac{1} {N} \sum\sum \omega_{ij}^2} = O(h_N) , $$ 
$$(iii) \;\;  \;\; \frac{1} {N} \sum\sum |\omega_{ij}| = O(h_N) . $$ 

\vspace{0.2cm}

\noindent{\bf Strong Cross-sectional Dependence:} Dependence across individuals is said to be strong when either of these three following conditions holds:
$$(i) \;\; \max_{i}(\sum_{j=1}^{N}|\omega_{ij}|) = O(N) , $$ 
$$(ii) \;\;  \;\; \sqrt{\frac{1} {N} \sum\sum \omega_{ij}^2} = O(N) , $$ 
$$(iii) \;\;  \;\; \frac{1} {N} \sum\sum |\omega_{ij}| = O(N) . $$ 

\vspace{0.2cm} 

It may be worthy to discuss the relationship among these three norms along with the maximum eigenvalue norm at this time. 

\vspace{0.2cm} 

\noindent{\it Proposition 2:} {\it $\sqrt {\frac{1} {N} \sum\sum \omega_{ij}^2} \le  \lambda_N \le \max_{i}(\sum_{j=1}^{N}|\omega_{ij}|). $ 
\footnote{ We do not have any such result for {\it taxicab} norm, in general.}  }

\vspace{0.5cm}

\noindent{\sc Proof:} Note that 
$ \sum\sum \omega_{ij}^2=Tr(\Omega \Omega')= \sum eigenvalue_i (\Omega \Omega')= \sum \lambda_i^2.$
Therefore, $ \frac{1} {N} \sum\sum \omega_{ij}^2= \frac{1} {N} \sum \lambda_i^2 \le \lambda_N^2.$  
Hence the first inequality. The second inequality i.e., $\lambda_N \le \max_{i}(\sum_{j=1}^{N}|\omega_{ij}|) $ 
may be found in Lewis(1991, Corollary 4.4.2). \qed

\vspace{0.2cm}

Equivalence holds except for few pathological but interesting  cases. Examples may be counted as below:

\vspace{0.2cm} 

\noindent{\it Example 12:} Consider a matrix whose one row and one column, say, the first one is such that all the 
elements except the diagonals  are same as $\frac{1} {c\sqrt{N}}, c\ge 2$. Elements of all other rows and columns 
are zeros except the diagonal elements. Whereas, all the diagonal elements are a fixed constant. Clearly,  
the maximum absolute row sum norm is unbounded and $O(\sqrt{ N})$ but the maximum eigenvalue is finite. Therefore, 
the maximum absolute row sum norm implies the dependence as moderate whereas weak dependence is implied by maximum 
eigenvalue norm. 

\vspace{0.2cm} 
Note that the above  example is not difficult to construct. For example, let $\{\varepsilon_t \} $ is a sequence  of i.i.d random variables with finite second moments.  Construct a sequence of new random variables  as $Y_1=\sum_{i=1}^n a_i \varepsilon_i , \; Y_j=b_j  \varepsilon_j, j=1 ,2 \ldots, n.$ Note that  $Cov(Y_j, Y_1) \ne 0, \; Cov(Y_j, Y_k)=0 \; \forall j \ne k \ne1.$

\vspace{0.2cm} 
This  example may look pathological in nature. However, this  example show that maximum eigenvalues norm is more logical and appropriate. 

\vspace{0.2cm} 

\noindent{\it Example 13:} Consider a matrix whose all the diagonal elements are same (say, $a<\infty$) and all the off-diagonal elements are also same (say, $b, 0 < b< a <\infty $). It is easy to see that maximum eigenvalue norm implies strong dependence. On the other hand,  the euclidean norm suggests dependence as moderate.

\vspace{0.2cm} 

\noindent{\it Example 14:} Consider a matrix whose all the diagonal elements are same (say, $a<\infty$) and all the off-diagonal elements are also same (say, $\frac{1} {\sqrt{N}} $). It is easy to see that maximum eigenvalue norm implies moderate dependence. On the other hand,  the euclidean norm suggests dependence as weak.

\vspace{0.5cm} 

\noindent{\it Proposition 3:} { \it  For $\lambda_N=O(N)$ if and only if 
$ \frac{1} {N} \sum\sum |\omega_{ij}| = O(N)$  if and only if $\sqrt{\sum\sum \omega_{ij}^2} = O(N).$ }
\vspace{0.2cm}

\noindent{\sc Proof:} \ (i) \ Suppose $\lambda_N=O(N)$. 
Since $\frac{1} {N} \sum\sum |\omega_{ij}| \le  C_0 N$ and
 $\sqrt{\sum\sum \omega_{ij}^2}  \le C_1 N$ for some positive constants $C_0, \ C_1$ hold always, it suffices
to prove the lower bound is also of the same order. 
By Jensen's inequality $\frac{1} {N^2} \sum\sum |\omega_{ij}| \le 
\sqrt{\frac{1} {N^2} \sum\sum \omega_{ij}^2}$
which is equivalent to
$\frac{1} {N} \sum\sum |\omega_{ij}| \le \sqrt{\sum\sum \omega_{ij}^2}$. But
 $O(N^2) \le \sum \lambda_i^2 = \sum\sum \omega_{ij}^2 \le O(N^2).$ Hence 
$\sqrt{\sum\sum \omega_{ij}^2} = O(N)$. 
Again, 
\begin{eqnarray*}
\label{cor}
\sqrt{\sum\sum \omega_{ij}^2} 
&\le & \max_i \omega_{ii} \sqrt{\sum\sum \omega_{ij}^2 /(\omega_{ii}\omega_{jj})} \nn\\
&\le & \max_i \omega_{ii} \sqrt{\sum\sum |\omega_{ij}|/\sqrt{\omega_{ii}\omega_{jj}}}  \nn\\
&\le & (\max_i \omega_{ii}/\sqrt{\min_i \omega_{ii}}) \sqrt{\sum\sum |\omega_{ij}|} . 
\end{eqnarray*}
Note that the second inequality is due to the fact that $ \frac{\omega_{i j}^2} {\omega_{i i}\omega_{j j}} \le 1.$
Thus, $\sqrt{\sum\sum |\omega_{ij}|} = O(N)$ This implies $\sum\sum |\omega_{ij}| = O(N^2)$
and hence the result, $(1/N) \sum\sum |\omega_{ij}| = O(N)$. Thus,
 $\lambda_N=O(N)$ implies $\sqrt{\sum\sum \omega_{ij}^2} = O(N)$ and $(1/N) \sum\sum |\omega_{ij}| = O(N)$.

\smallskip
(ii) \ For other way,
let $(1/N) \sum\sum |\omega_{ij}| = O(N)$, then by
Jensen's inequality $\sqrt{\sum\sum \omega_{ij}^2} \ge O(N)$.
But $\sum\sum \omega_{ij}^2 \le O(N^2)$. Hence $\sqrt{\sum\sum \omega_{ij}^2} = O(N)$.
Again, $\sum\sum \omega_{ij}^2 = \sum \lambda_i^2$. Hence $\sum \lambda_i^2 = O(N^2).$

Note that $\sum \lambda_i = \sum \omega_{ii} = O(N)$. This is due to the following fact:
\begin{eqnarray*}
{\sum\sum \omega_{ij}^2} 
&\le & \sum\sum (\frac{\omega_{ij}^2} {\omega_{ii}\omega_{jj}}) \omega_{ii} \omega_{jj} \nn\\
&\le & \sum\sum \omega_{ii} \omega_{jj}  \nn\\
&=& ( \sum \omega_{ii})^2.
\end{eqnarray*}
This implies that $\sum \omega_{ii} \ge O(N)$ which in turn implies $\sum \omega_{ii} = O(N), $  as $ \sum \omega_{ii} \le O(N).$
So, if possible, assume the largest eigenvalue
is of lower order than $N$, say $h_1(N)$, such that $h_1(N)/N \to 0$ as $N \to \infty$. If 
$h_1(N)$ is of lower order than $(\sqrt{N})$, then $\sum \lambda_i^2/N^2 \le N (h_1(N))^2/N^2 \to 0$,
as $N \to \infty$, 
so contradiction. Hence, $h_1(N)$ must be at least of order $\sqrt{N}$. Let there be $g_1(N)$ many
eigenvalue of order $h_1(N)$. Then $g_1(N) \le O(N/h_1(N))$, as $g_1(N) h_1(N) \le \sum \lambda_i = O(N)$.
Thus $g_1(N) (h_1(N))^2/N^2 \le O(N/h_1(N)) (h_1(N))^2/N^2  = O(N h_1(N))/N^2 \to 0$ as $N \to \infty$. 
Hence there must
be another eigenvalue of order $h_2(N)$ such that $h_2(N)/h_1(N) \to 0$ as $N \to \infty$.
 Let there be $g_2(N)$ many
eigenvalue of order $h_2(N)$. Then $g_2(N) \le O((N - g_1(N) h_1(N))/h_2(N))$, as 
$g_2(N) h_2(N) \le \sum_{i \le N-g_1(N)} \lambda_i = O(N - g_1(N) h_1(N))$.
Continue like these till one exhausts all $N$ eigenvalues. 
Thus, one obtains $\sum g_j(N) h_j(N) \le \sum \lambda_i = O(N)$.
Now, 
\begin{eqnarray*}
\bigl(\sum \lambda_i^2\bigr)/N^2 
&=& \bigl(\sum g_j(N) (h_j(N))^2\bigr)/N^2  \nn\\
& \le & h_1(N) \bigl(\sum g_j(N) (h_j(N))\bigr)/N^2 \nn\\
&=& \bigl(h_1(N) (O(N))\bigr)/N^2  \nn\\
&\rightarrow & 0 \ \ \ \mbox{  as   } N \to \infty .
\end{eqnarray*}
Thus, contradiction! Hence the order of the largest eigenvalue $h_1(N)$ must be $N$, i.e., $\lambda_N = O(N)$.
Therefore,
 $(1/N) \sum\sum |\omega_{ij}| = O(N)$ implies $\sqrt{\sum\sum \omega_{ij}^2} = O(N)$ and $\lambda_N=O(N)$.

\smallskip
(iii) \ For the last part, assume that $\sqrt{\sum\sum \omega_{ij}^2} = O(N)$,
then by above argument (i),
$\sqrt{\sum\sum \omega_{ij}^2} \le (\max_i \omega_{ii}/\sqrt{\min_i \omega_{ii}}) \sqrt{\sum\sum |\omega_{ij}|}$
and hence  
$\sqrt{\sum\sum |\omega_{ij}|} \ge O(N)$, i.e., 
$\sum\sum |\omega_{ij}| \ge O(N^2)$, and hence
$(1/N) \sum\sum |\omega_{ij}| \ge O(N)$. Since
$(1/N) \sum\sum |\omega_{ij}| \le O(N)$, it implies,
$(1/N) \sum\sum |\omega_{ij}| = O(N)$. 

Again, from the above argument in (ii), $\sum \lambda_i = \sum \omega_{ii} = O(N)$ and 
$\sum \lambda_i^2 = \sum\sum \omega_{ij}^2 = O(N^2)$, imply
that the largest eigenvalue $\lambda_N$ must be of order $N$. 
Thus,
$\sqrt{\sum\sum \omega_{ij}^2} = O(N)$ implies that  $(1/N) \sum\sum |\omega_{ij}| = O(N)$
 and $\lambda_N=O(N)$.

\medskip
Hence the proof of Proposition 3 is complete.
\qed

\bigskip
{\bf Conjecture:} { \it  For $\lambda_N=O(1)$ if and only if 
$ \frac{1} {N} \sum\sum |\omega_{ij}| = O(1)$  if and only if $\sqrt{\sum\sum \omega_{ij}^2} = O(\sqrt{N}).$}

\vspace{0.5cm} 

\section{Linear Panel Data  Models}
Here we will consider two popularly used linear models, viz., fixed effect model and random effect (pooled) model.  All these models are considered under one-way error component structure. Here we intend to derive asymptotic properties  of two omnipresent  parameter estimators, viz., the pooled estimator and the fixed effect estimator. We consider two different situations separately: viz., (1) only cross-sectional dependence, no time dependence; and (2) both cross-sectional and time series dependence. Separate treatment is important to get more insight. 

\subsection{ Only Cross-Sectional Dependence, No Time Dependence: Fixed and Random Effect Models} 

Let $y_{it}$ be the observation on the $i$-th cross-section unit at time $t$ for $i =
1, 2, ..., N, t = 1, 2,..., T,$ and suppose that it is generated according to the
linear heterogeneous panel data model as
\begin{equation}
y_{it}=\mu _{i}+ x_{i,t}' \beta + \varepsilon _{it},   
\end{equation}
where  $x_{it}$ is a $k \times 1$ vector of observed
individual-specific regressors on the $i$-th cross-section unit at time $t$.
  
 Defining $Y_{i .}=(y_{i1}, y_{i2}, \ldots, y_{iT})', X_{i .}= (x_{i1}, x_{i2}, \ldots, x_{iT})', 1_T=(1,1,\ldots ,1)'$, and  $\varepsilon_{i .}= (\varepsilon_{i1},\ldots,\varepsilon_{iT})'$ we can write (1) in matrix notation as 
 \begin{equation}
Y_{i .}=\mu _{i} 1_T+X_{i .} \beta+\varepsilon _{i .}.   
\end{equation}
 Further, defining $Y=(Y_{1 .}', Y_{2 .}',\ldots , Y_{N .}')',  X=(X_{1 .}', X_{2 .}',\ldots , X_{N .}')', \varepsilon= (\varepsilon_{1 .}',\ldots,\varepsilon_{N .}')'$,  $\mu=(\mu_1, \mu_2,\ldots, \mu_N)'$ and $D=I_N \otimes 1_T $ we can rewrite (2) in matrix notation as 
  \begin{equation}
Y=D \mu + X \beta+\varepsilon,   
\end{equation}
where $I_N$ is an identity matrix and $ \otimes$ denote the Kronecker product. 

Another matrix form of the equation (1) by stacking $N$ observations for each time point instead of  the above form may be helpful for our purpose.

 Define $Y_{. t}=(y_{1t}, y_{2t}, \ldots, y_{Nt})', X_{. t}= (x_{1t}, x_{2t}, \ldots, x_{Nt})'$, and  $\varepsilon_{. t}= (\varepsilon_{1t},\ldots,\varepsilon_{Nt})'$  and write (1) in matrix notation as 
 \begin{equation}
Y_{. t}=\mu +X_{. t} \beta+\varepsilon _{. t}.   
\end{equation}

 Further, defining $ \mathscr{Y}=(Y_{. 1}', Y_{. 2}',\ldots , Y_{. T}')',  \mathscr{X}=(X_{. 1}', X_{. 2}',\ldots , X_{. T}')', \mathscr{E}= (\varepsilon_{. 1}',\ldots,\varepsilon_{. T}')'$,   and $ \mathscr{D}=1_T  \otimes I_N $ we can rewrite (4) in matrix notation as 
  \begin{equation}
\mathscr{Y}=\mathscr{D} \mu + \mathscr{X} \beta+ \mathscr{E}.   
\end{equation}

Consider now the following assumptions \footnote {We need a few more assumptions for CLT. We will state these assumptions at the appropriate place.}.  We will use two norms,viz., maximum eigenvalue norm and the trace norm. The  maximum eigenvalue norm for any positive definite matrix is defined as $ \displaystyle ||A||_{e}=  \max_{l}{l'Al}$.  The trace norm is defined as $||A||_{s}= [tr(A'A)]^{1/2}$.

\vspace{0.2cm}

\noindent{\bf Assumption A1:}  We assume that  $\frac {X'X}{NT}$ converges to  $Q$ in mean,  where $ Q= plim_{N,T} (\frac {X'X}{NT})$ is a finite and non-singular matrix. 

We also assume that $\frac {1} {NT} \sum \sum X_{i t} $  converges in mean square. Furthermore, we assume that both the matrices, $\frac {X'MX}{NT}$ and $\frac {X'\bar{M} X}{NT}$ converge to finite nonsingular matrices in
probability. Also,  $\frac {X_{.t}\mathscr{M}_t' \mathscr{M}_t X_{.t}'}{N}$ converges to a finite and non-singular matrix.The matrices $\mathscr{M}_t$, $M$ and $\bar{M}$ are defined below.

\vspace{0.2cm}

\noindent{\bf Assumption A2:}  We further assume that $\mu_i$ and $ x_{i1}, x_{i2}, \ldots, x_{iT}$ are strictly 
exogenous with respect to $\varepsilon_{it}$; that is, 
 $$ E(\varepsilon_{it}| \mu_i,  x_{i1}, x_{i2}, \ldots, x_{iT})=0,$$
for any $i$ and $t$. This assumption rules out the possibility of inclusion of any lag dependent variables or any 
predetermined regressors. Though this strict exogeneity assumption is overly restrictive, it facilitates 
the derivations greatly. However, this kind of assumption is not uncommon in the literature, see for example, 
Gon\c{c}alves (2011). 

\vspace{0.2cm}

\noindent{\bf Assumption A3:} The unobserved heterogeneity parameter $\mu_i$ is assumed to be bounded.

\vspace{0.2cm}

\noindent{\bf Assumption A4:}   $ \sup_{i,t} E|\varepsilon_{i t}|^{2r} < \infty, $ for some  $r\ge 2,$ need not be an integer. 

\vspace{0.2cm}


\vspace{0.2cm}

\noindent{\bf Assumption A5:}    $ \varepsilon_{. t}$ is i.i.d. \footnote{ The i.i.d. assumption is for simplicity. In Section 3.2, we will allow various forms of time  dependence.}   with $E\ \varepsilon_{. t}=0$ and with variance   $\Omega .$

\vspace{0.2cm}

Depending upon whether $E(\mu_i|X_{it})$ is zero or not, we have two known models; random effect and fixed effect models, respectively. For fixed  effect model we consider the fixed effect or the within estimator. On the other hand, OLS estimator as opposed to GLS estimator is considered for random effect model.

\vspace{0.4cm}

We consider two popular estimators of $\beta$  as:

\vspace{0.2cm}

$\hat{\beta}_{FE} =(X'MX)^{-1}X'MY \equiv (\mathscr{X}'\mathscr{M}\mathscr{X})^{-1} \mathscr{X}'\mathscr{M}\mathscr{Y} $ as $(X'MX)= (\mathscr{X}'\mathscr{M}\mathscr{X})$  and $ X'MY=\mathscr{X}'\mathscr{M}\mathscr{Y}, $ \\

 where, \;  $M=I_{NT}-D(D'D)^{-1}D', \;\; D =I_N \otimes 1_T, $ \;   and 
 
  \; $ \mathscr{M}=I_{NT}-\mathscr{D}(\mathscr{D}'\mathscr{D})^{-1}\mathscr{D}', \;\; \mathscr{D}=1_T \otimes I_N. $ 

\vspace{0.2cm}

Similarly,

$ \hat{\beta}_{ols} =(X'\bar{M}X)^{-1} X'\bar{M}Y,$ \; where, \;  $\bar{M}=I_{NT}-\bar{D}(\bar{D}'\bar{D})^{-1}\bar{D}', \;\; \bar{D}=1_{NT}.$
We can define $\mathscr{\bar{M}}$ similar to $\mathscr{M}$ by defining $\mathscr{\bar{D}}=\bar{D}$. Thus, it is easy to see that $\mathscr{\bar{M}}= \bar{M}.$

\vspace{0.4cm}

$\hat{\beta}_{FE}$ may be written as
\[\hat{\beta}_{FE}= (X'MX)^{-1} \sum_{t=1}^{T} B_t Y_{.t},
\]
where  $B_t=\mathscr{X}' \mathscr{M}_t $ \;, where, \;  $\mathscr{M}_t $ is the 
$t$-th block of $ \mathscr{M}=[\mathscr{M}_1 \vdots \mathscr{M}_2 \vdots \; \cdots \;  \vdots \mathscr{M}_T].$

Here it may be noted that $ \mathscr{M}=  \mathscr{M}  \mathscr{M}'= \sum \mathscr{M}_t \mathscr{M}_t'$ . 
It is also easy to see that  $ \mathscr{M}=  \mathscr{M} \mathscr{\bar{M}}$ i.e., $\sum \mathscr{M}_t \mathscr{\bar{M}}_t'= \mathscr{M}  \mathscr{\bar{M}}= \mathscr{M}$. 

\vspace{0.4cm}

Similarly, $ \hat{\beta}_{ols}$ may be written as
\[ \hat{\beta}_{ols}= (X'\bar{M}X)^{-1} \sum_{t=1} ^ {T} C_t Y_{.t} ,
\]
 where  $C_t=(X_{.t}'- [\frac {X' 1_{NT}1_{N}'}{1_{NT}'1_{NT}}])$.
 
 Define $b_t=(X'MX)^{-1}B_t$ and $c_t=(X'\bar{M}X)^{-1}C_t. $

\vspace{0.4cm}

To visualize the elements of $b_t$, consider  a single regressor( for simplicity). In the single regressor case, 
the $i$-th element of $b_t$ looks like $b_{it}=\frac{X_{it}-\Bar{X_i}}{\sum \sum (X_{it}-\Bar{X_i})^2}.$ Similarly 
$c_{it}= \frac{X_{it}-\Bar{\Bar{X}}}{\sum \sum (X_{it}-\Bar{\Bar{X}})^2}$.
It is easy to see that $E(b_t b_t')=O(\frac{1}{NT^2}).$  This is also true that $E(c_t c_t')=O(\frac{1}{NT^2}).$ 

\vspace{0.4cm}

Let $V_{FE}=Var_X(\hat{\beta}_{FE})=(X'MX)^{-1}(X'M'\Omega M X)(X'MX)^{-1}.$

Let $V_{ols}=Var_X(\hat{\beta}_{ols})=(X'\bar{M}X)^{-1}(X'\bar{M}'\Omega \bar{M} X)(X'\bar{M}X)^{-1}.$

\vspace{0.4cm}

Following theorem provides the properties of these two estimators.

\vspace{0.2cm}

\noindent{\bf Theorem 1:}  {\it Under all the assumptions stated above, from {\bf  A1} to {\bf  A5},  both the estimators, 
$\hat{\beta}_{FE} $ and  $ \hat{\beta}_{ols}$ are consistent.} 

\vspace{0.4cm}
\noindent{\bf Remark 2:} The proof of this theorem shows that,  $\hat{\beta}_{FE} $ and  $ \hat{\beta}_{ols}$ are $\sqrt {T} $ 
consistent   when  the dependence is strong and  are $\sqrt {NT} $ consistent for weak dependence. It may be interesting to 
examine consistency of the estimators when $T$ is finite, say $T=1.$ It is easy to see that both $\hat{\beta}_{FE} $ and  
$ \hat{\beta}_{ols}$ are inconsistent when dependence is strong. Even though error-factors do not have any relationship with 
the included regressors, both the estimators are inconsistent, in general. 

\vspace{0.4cm}
\noindent{\bf Remark 3:} In some situations, even when $T$ is finite,  $\hat{\beta}_{FE} $ and  $ \hat{\beta}_{ols}$ are 
consistent under strong dependence. For simplicity, consider the case when $T=1$ and consider  $\hat{\beta}_{ols}.$  Let 
us concentrate on the term 
$(X'\Omega X)=\sum_{i=1}^{N} \lambda_i Z_i Z_i' =\sum_{i=1}^{N-m} \lambda_i Z_i Z_i' +\sum_{i=N-m+1}^{N} \lambda_i Z_i Z_i', $ 
where $Z_i=P_i X$, $P$ is the matrix of eigenvectors coming from the spectral decomposition of $\Omega.$  Assume, as earlier, 
there are $m$ (finite) eigenvalues; $\lambda_N \ge \lambda_{N-1} \ge, \ldots, \lambda_{N-m+1},$ which are    of $O(N).$  
Note that, $ \lim \frac{1}{N} \sum_{i=1}^{N}  Z_i Z_i'=\lim\frac{X'X}{N},  $ is a finite non-singular matrix. Now {\bf assume} 
that  $ lim \frac{1}{N} \sum_{i=N-m+1}^{N}  Z_i Z_i' \rightarrow 0. $ This can happen when either $P_i$ is orthogonal to $X$, 
or $P_iX=o(\sqrt{N}).$  Under this assumption, it is easy to see that both the terms, 
$\frac{1}{N^2}\sum_{i=1}^{N-m} \lambda_i Z_i Z_i'$ and 
$ \frac{1}{N^2}\sum_{i=N-m+1}^{N} \lambda_i Z_i Z_i'  \rightarrow 0 \; as \; N \rightarrow \infty.$  Hence, 
$ Var(\hat{\beta}_{ols})=(X'X)^{-1}(X'\Omega X)(X'X)^{-1}
= (\frac{X'X}{N})^{-1}(\frac{X'\Omega X}{N^2})(\frac{X'X}{N})^{-1}\rightarrow 0 \; as \; N \rightarrow \infty.$  
This observation can be illustrated by a simple example.  Consider a simple linear regression model as 
$y_i=\beta X_i +\epsilon_i,\;\; i=1, 2, \ldots, N$ with $Var(\epsilon_i)=a,\; \forall \; i=1,2, \ldots, N$ and 
$Cov(\epsilon_i, \epsilon_j)=b (>0)\; \forall \; i\ne j =1,2, \ldots, N. $ Assume all other standard assumptions hold. 
It is straight forward to see that $ \hat{\beta}_{ols}$ is inconsistent, in general. However, $ \hat{\beta}_{ols}$ is 
consistent when $\sum_{i=1}^{N} X_i =0,$ or  $O(1),$ or $O(h_N), \frac{h_N}{N}\rightarrow 0,$ since the  eigenvector 
corresponding to the largest eigenvalue is the normalized form of $(1,1,\ldots, 1).$  One interesting real-life situation
is when $X$ is centered at $0.$

\vspace{0.2cm} 

The following theorem provides the asymptotic distribution of the estimators:

\vspace{0.2cm}
\noindent{\bf Theorem 2:}  {\it Consider the model in (1) along with all the five assumptions. 
Further assume that,
(i) for strong dependence,
an eigenvector  corresponding to the 
largest eigenvalue of $\Omega$ belongs to the row-space of $(\sum b_t' b_t)$; or $(\sum c_t' c_t),$ for fixed and random effect model, respectively
and 
(ii) for weak dependence, the largest eigenvalue of
$E[(\varepsilon_t \varepsilon_t') \otimes (\varepsilon_t \varepsilon_t')]$ is of order
$O(N)$.
  
Then (a) $V_{FE}^{-\frac{1}{2}}(\hat{\beta}_{FE} -\beta) \to N(0, I_K)$ (i) as $ \frac{1}{T} \to 0,  $ irrspective of any (large) $N$ for strong dependence; and (ii) as 
$ \frac{ N} { T} \to 0,  $  for weak dependence.  (b) $V_{ols}^{-\frac{1}{2}}(\hat{\beta}_{ols} -\beta) \to N(0, I_K)$  (i) as $ \frac{1}{T} \to 0,  $ irrspective of any (large) $N$ for strong dependence; and (ii) as  
$ \frac{ N} { T} \to 0,  $  for weak dependence. } 

\vspace{0.5cm}

Different rate of convergence to asymptotic normality across various dependence may be possible under several situations. Let us 
consider one such situation.  Let $\Omega_c$ is a matrix whose $(i,\;j)-$th element is 
$E(\varepsilon_{it}\varepsilon_{jt}|\varepsilon_{kt}\varepsilon_{lt}),\; k,l\ne i,j.$ 
Assume that $(l' \Omega_c l) \le C\; l'l\; \lambda_N=C\lambda_N,\; C$  is independent of $k,l, $  and assuming $l'l=1.$ 
This assumption holds, for example, for normal distribution of $\varepsilon_{it}.$ Then it is easy to check that 
$\sum E_X(l' b_t \varepsilon_t \varepsilon_t' b_t' l) ^2 \le (\frac {1}{NT^2})^2 \; C^2 \;  \lambda_N^2 \; T.$ Hence 
$ \frac{\sum E_X(l' b_t \varepsilon_t \varepsilon_t' b_t' l) ^2  } { ( \sum l' b_t \Omega b_t' l) ^2 } 
=  O (\frac{\lambda_N^2} {T}). $

\vspace{0.2cm}

The following corollary provides the asymptotic distribution of the estimators under the above situation:

\vspace{0.5cm}

\noindent{\bf Corollary 1:} {\it Consider  the above situation along with all the assumptions as in Theorem 2. Then  
(a) $V_{FE}^{-\frac{1}{2}}(\hat{\beta}_{FE} -\beta) \to N(0, I_K)$ as $ \frac{N^2} {T} \to 0 $, under strong dependence, 
(b) $V_{FE}^{-\frac{1}{2}}(\hat{\beta}_{FE} -\beta) \to N(0, I_K)$ as $ \frac{h_N^2} {T} \to 0 $, under semi-strong dependence, 
and (c) $V_{FE}^{-\frac{1}{2}}(\hat{\beta}_{FE} -\beta) \to N(0, I_K)$ as $ \frac{N} {T} \to 0 $, under weak  dependence. 
Exactly similar results hold for  $V_{ols}^{-\frac{1}{2}}(\hat{\beta}_{ols} -\beta)$.}  

\vspace{0.5cm}

\vspace{0.5cm}

\noindent{\bf Remark 4:} Unconditional central limit theorems hold under the additional assumption of  
$ \frac{\sum (l' b_t \Omega b_t' l)   } { E( \sum l' b_t \Omega b_t' l)  } -1 = o_p(1), \; \forall l \in R^k $  
such that $l'l=1.$  This can also be achieved if the convergence of $\sum ( b_t \Omega b_t' )$  to the 
$E( \sum  b_t \Omega b_t')$ holds in probability with appropriate scaling of $N$ and $T$. Similarly, unconditional CLT can be 
derived for $\hat{\beta}_{ols}$ by replacing $b_t$ by $c_t.$

\vspace{0.5cm}

\noindent{\bf Remark 5:} Consider  $\sum E_X(l' b_t \varepsilon_t \varepsilon_t' b_t' l) ^r, \; r>1,  $ instead of 
$\sum E_X(l' b_t \varepsilon_t \varepsilon_t' b_t' l) ^2. $ Then it is easy to see that the theorem holds as  
$ \frac{N^r} {T^{r-1}} \to 0 .$ It is further to see that for $\varepsilon_{it}$   with MGF,  Theorem 2 holds when  
$ \frac{N} {T} \rightarrow 0. $

\vspace{0.2cm}
\bigskip

\noindent{\bf Remark 6:}
Define a matrix based on fourth moments and cross moments whose dimension is $N^2 \times N^2$.
It is given by $V_F = E((\varepsilon_t \varepsilon_t') \otimes (\varepsilon_t \varepsilon_t'))$.
Note, $trace(V_F) \ge \lambda_{\max}(V_F) \ge  \sum_{i,j}\omega^2_{ij} 
\ge \lambda_{\max} (\Omega^2) = \lambda_{\max}^2 (\Omega) $.
Hence for strong dependence $O(\lambda_{\max}(V_F)) = O(N^2)$, 
whereas, for weak dependence, in general, $O(\lambda_{\max}(V_F)) \ge O(N)$. Under this condition, Theorem 2 holds when 
$ \frac{N} {T} \rightarrow 0. $

\vspace{0.2cm}
\bigskip

\noindent{\bf Remark 7:} For strong dependence with $\lambda_{min} (\Omega)=O(1)$, however, we may be able to achieve the rate 
of $ (\frac{N^{2(1-s)}} { T})$ for some $0 \le s \le 1.$ This is because $ Z'\Lambda \Lambda' Z / ||Z||^2 $ is $O(N)$ whenever 
$Z$ (an $N$ dimensional  vector) $\in col(\Lambda),$ column space of $\Lambda,$  and $ Z'\Lambda  =0 $ if $Z \in null (\Lambda'), $ 
the null space of $\Lambda',$ which is orthogonal to the column space of $\Lambda.$  Every vector $Z$ can be decomposed into two 
parts $Z_1$ and $Z_2$  such that $Z_1 \in   col(\Lambda)$, and $Z_2 \in null (\Lambda').$  Hence,  $ Z'\Lambda \Lambda' Z / ||Z||^2 
= Z_1'\Lambda \Lambda' Z_1 / ||Z||^2 = (Z_1'\Lambda \Lambda' Z_1 / ||Z_1||^2) (||Z_1||^2/||Z||^2) = O(N)(||Z_1||^2/||Z||^2),$  
which is of $O(N^{s})$, for some $ s<1.$ Since, for $T>N$, with probability 1, one expects $b_t$ not to lie in the column space of 
$\Lambda$ (of fixed dimension $m$) for all $t$. Hence, $  \sum (l' b_t \Lambda \Lambda' b_t' l)$ should be $O(\frac{N^s}{NT}),$ 
which implies $  \sum (l' b_t \Omega b_t' l)= O(\frac{N^s}{NT})$. Therefore, one expects to get the  rate of 
$ (\frac{N^{2(1-s)}} {T}). $

\vspace{0.5cm}

\subsubsection{Estimation of Variance Under Pure Cross-sectional: Fixed and Random Effect Models}

It is to note that the above asymptotic distributions are not implementable in practice as the matrix $V_{FE}$ and $V_{ols}$ are 
unknown. For fixed $X$, one needs only the estimate of $\Omega$.    One natural estimator of $\Omega$ is 
 \[
\widehat \Omega = \frac{1}{T} \sum_{t=1}^T e_t e_t',
\]
where $ e_t$ is the estimated residual obtained by using the corresponding residuals.  This estimator of $\Omega$ is quite general 
in nature. For fixed $N$, it is not difficult to prove consistency of  $ \widehat \Omega $. Therefore, for fixed $N$, natural 
estimator of $V_{FE}$ or  $V_{ols}$ may be obtained by replacing $\Omega$ by  $\widehat \Omega.$

The problem with this estimator is that, even when $T$ is sufficiently large
relative to $N$ for the estimator to be feasible, its finite sample properties may be quite poor in situations where $N$ and $T$ 
are of comparable orders of magnitude. This is so because the many elements of the cross-sectional covariance matrix will be 
poorly estimated.

In this paper we propose a simple modification of the
 covariance matrix estimator
which remedies the deficiencies of techniques which rely
on large $T$ asymptotic. A closer look into the construction of $V_{FE}$ and $V_{ols}$  suggest that both the variances are  
$k \times k $ matrices. So, at least theoretically,  it is possible to estimate $V_{FE}$ and $V_{ols}$ without the  restriction 
of fixed $N$. Moreover, it is  invertible for any combination of $N$ and $T$, including the case when $N>T$. 

We propose an estimator of $V_{FE}$  as
\[
\bar{V}_{FE} =  \sum b_t e_t e_t' b_t'.
\]

Similarly, we propose an estimator of $V_{ols}$ as
\[
\bar{V}_{ols} =  \sum c_t e_t e_t' c_t'.
\]

Here it may be noted that \[ b_t=(X'MX)^{-1}B_t,
\]
and
\[c_t= (X'\bar{M}X)^{-1}C_t.
\]

\vspace{0.4cm}

Following theorem provides the properties of these two estimators.

\vspace{0.4cm}

\noindent{\bf Theorem 3:}  {\it Under all the assumptions stated above, from {\bf  A1} to {\bf  A5},  
both the estimators, $\bar{ V}_{FE} $ and  $\bar{V}_{ols} $  are consistent as $T \to \infty $.}

The following theorem provides the asymptotic distribution of the parameter estimators with estimated variance covariance matrices:

\vspace{0.2cm}
\noindent{\bf Theorem 4:}  {\it Consider the model in (1) along with all the five assumptions. 
Further assume that,
(i) for strong dependence,
an eigenvector  corresponding to the 
largest eigenvalue of $\Omega$ belongs to the row-space of $(\sum b_t' b_t)$; or $(\sum c_t' c_t),$ for fixed and random effect model, respectively
and 
(ii) for weak dependence, the largest eigenvalue of
$E[(\varepsilon_t \varepsilon_t') \otimes (\varepsilon_t \varepsilon_t')]$ is of order
$O(N)$.
  
Then (a) $\bar{V}_{FE}^{-\frac{1}{2}}(\hat{\beta}_{FE} -\beta) \to N(0, I_K)$ (i) as $ \frac{1}{T} \to 0,  $ irrespective of any (large) $N$ for strong dependence; and (ii) as  
$ \frac{ N} { T} \to 0,  $  for weak dependence.  (b) $\bar{V}_{ols}^{-\frac{1}{2}}(\hat{\beta}_{ols} -\beta) \to N(0, I_K)$  (i) as $ \frac{1}{T} \to 0,  $ irrespective of any (large) $N$ for strong dependence; and (ii) as  
$ \frac{ N} { T} \to 0,  $  for weak dependence. } 

\vspace{0.5cm}

Consider testing linear hypotheses about $\beta$ of the form
$H_0: R\beta = r$,
where $R$ is a $q\times k$ matrix of known constants with full rank with
$q \le k$ and $r$ is a $q \times 1$ vector of known constants. Define the Wald
statistics as
\[
Wald_{FE} = (R \Hat{\beta}_{FE}-R\beta)^{\prime} [R\bar{V}_{FE}R']^{-1} (R \Hat{\beta}_{FE}-R \beta), \]

and

\[
Wald_{ols} = (R \Hat{\beta}_{ols}-R\beta)^{\prime} [R\bar{V}_{ols}R']^{-1} (R \Hat{\beta}_{ols}-R\beta). \]

The following theorem provides the distribution of the above wald statistics under the null hypothesis.

\vspace{0.2cm}
\noindent{\bf Theorem 5:}  {\it Consider the model in (1) along with all the five assumptions. 
Further assume that,
(i) for strong dependence,
an eigenvector  corresponding to the 
largest eigenvalue of $\Omega$ belongs to the row-space of $(\sum b_t' b_t)$; or $(\sum c_t' c_t),$ for fixed and random effect model, respectively
and 
(ii) for weak dependence, the largest eigenvalue of
$E[(\varepsilon_t \varepsilon_t') \otimes (\varepsilon_t \varepsilon_t')]$ is of order
$O(N)$.
  
Then (a) $Wald_{FE} \to \chi_k^2 $ (i) as $ \frac{1}{T} \to 0,  $ irrespective of any (large) $N$ for strong dependence; and (ii) as  
$ \frac{ N} { T} \to 0,  $  for weak dependence.  (b) $Wald_{ols} \to \chi_k^2 $  (i) as $ \frac{1}{T} \to 0,  $ irrespective of any (large) $N$ for strong dependence; and (ii) as 
$ \frac{ N} { T} \to 0,  $  for weak dependence. }

\vspace{0.5cm}

\subsection{Both Cross-sectional and Time Series Dependence: Fixed and Random Effect Models}

{\it Theorem 1}, in the previous sub-section, considers consistency of both fixed effect and pooled estimator. Consistency is 
derived by considering both time and cross-sectional dependence.  So far asymptotic normality is derived only under the 
cross-sectional dependence without the time dependence. In this sub-section, we present asymptotic normality under various 
pragmatic situations. We consider three different kind of situations with error factor structure \footnote{ for brevity, pooled 
model is ignored}. The error factor structure is quite general to accommodate various kinds of dependence, viz,; strong, 
semi-strong, weak etc;  as it follows from the {\it Proposition 1} and the subsequent remark ({\it Remark 1 }). In this section, 
for simplicity,  for all kinds of dependence, viz,; strong, semi-strong, weak, we use a common notation of $h_{N}.$

Let $y_{it}$ be the observation on the $i$-th cross-section unit at time $t$ for $i =
1, 2, ..., N, t = 1, 2,..., T,$ and suppose that it is generated according to the
linear heterogeneous panel data model as
\begin{equation}
y_{it}=\mu _{i}+ x_{i,t}' \beta + \varepsilon _{it}, \;\; \\ 
\varepsilon _{it}=\lambda_i f_t +u_{it}, 
\end{equation}
where  $x_{it}$ is a $k \times 1$ vector of observed
individual-specific regressors on the $i$-th cross-section unit at time $t$.

In vector form, the above model is  $\varepsilon_{.t}= \Lambda f_t+ u_{.t} $, $\Lambda$ is  an $N\times m$ matrix of  
factor loadings with $ lim_{N \rightarrow \infty} (\frac{\Lambda' \Lambda} {h_N}),$  
\footnote {Either, $h_N=O(1)$; or $h_N \uparrow \infty$ as $N\uparrow \infty; $ whereas  for later case,   $h_N=o(N)$ or $h_N= O(N). $ } is a 
positive definite
matrix.  The  $ u_{.t}= [u_{1t},\ldots,u_{Nt}]'$ is an $N \times 1$ vector of  errors  with
$E(u_{.t})=0$ and $E(u_{.t} u_{.t}')=\Sigma.$    Note that 
$ \Omega =Cov (\varepsilon_{.t})=\Lambda  \Lambda ^{\prime} +\Sigma, $  \; where   
$  (\Lambda \Lambda^{\prime}) $ and $\Sigma$ both are non-negative definite matrices. The further assumptions on 
$ f_t, \; u_{it}$ will be made as we proceed \footnote{ for brevity, proofs of following theorems are ignored}.

\vspace{0.2cm}
\noindent{\bf CASE 1a:}  {\it Consider the model in (6) along with assumptions {\bf A1 to A4}.  Further assume that
$u_{it}$ does not have any cross-sectional dependence but have finite time memory (say of order $q$), 
 or covariance summability property. 
Also, assume $f_t$ is 
independent over time and  $h_N \to \infty.$ Then
$$ \sqrt\frac{NT}{h_N} (\hat{\beta}_{FE}-\beta) \rightarrow N(0, A_0),$$
where
$A_0=lim_{N,T} \sum b_t\Lambda \Lambda^{\prime} b_t^{\prime} \frac{NT}{h_N}.$ }

\vspace{0.5cm}

\noindent{\bf Remark 8:} Note that here cross-sectional dependence in $\varepsilon _{it}$ is either strong or semi-strong
and $Cov(\varepsilon_t, \varepsilon_{t-k})=Cov(u_t, u_{t-k})=\Omega_{k}$ for $k\neq 0$ and 
$Cov(\varepsilon_t, \varepsilon_{t}) = \Omega_{0} = \Lambda \Lambda^{\prime} + \Sigma$.
Closer look at $A_0$ 
shows that it does not 
involve $\Sigma$ and other cross-covariances of $u_t.$ It is due to the fact that,  
$lim_{N,T} \sum_t \sum_k b_t \Omega_k b_{t+k}^{\prime} \frac{NT}{h_N}
=lim_{N,T} \sum_t b_t\Lambda \Lambda^{\prime} b_t^{\prime} \frac{NT}{h_N}
+lim_{N,T} \sum_t b_t\Sigma b_{t}^{\prime} \frac{NT}{h_N}
+lim_{N,T} \sum_t \sum_{k > 0} b_t\Omega_k b_{t+k}^{\prime} \frac{NT}{h_N}.$ 
In the limit, the second and the third terms vanish. 

\noindent{\bf Note:} It is interesting to observe that time dependence in $\varepsilon _{it}$  coming through $u_{it}$  
does not have any influence in the asymptotic variance of $(\hat{\beta}_{FE}).$ Hence, in  estimating $Var(\hat{\beta}_{FE}),$ 
Newey-West kind of correction injects bias in the variance estimate. 

\vspace{0.2cm}
\noindent{\bf CASE 1b:}  {\it Consider the model in (6) along with assumptions {\bf A1 to A4}.  Further assume that 
$\varepsilon_{it}$ is having weak cross-sectional dependence but have finite time memory, say of order $q$.  Then
$$ \sqrt {NT} (\hat{\beta}_{FE}-\beta) \rightarrow N(0, A_1),$$
where
$A_1=lim_{N,T} (X'MX)^{-1} X'M' \Gamma MX (X'MX)^{-1} NT,$
and
$\Gamma=(\Gamma_{ij})$ \;
where \; $\Gamma_{ij} = \Omega_{k}=Cov(\varepsilon_t, \varepsilon_{t-k})$, if \ $|i-j| = k \le q$
and $\Gamma_{ij} = {\bf 0}$ matrix, if \ $|i-j| = k > q$. }

\vspace{0.2cm}
\noindent{\bf CASE 1c:}  {\it Consider the model in (6) along with assumptions {\bf A1 to A4}.  Further assume that 
$\varepsilon_{it}$ is having weak cross-sectional dependence but have ergodic time memory. In this case, when CLT holds for 
$\sum b_t \varepsilon_t$, then
$$ \sqrt {NT} (\hat{\beta}_{FE}-\beta) \rightarrow N(0, A_2),$$
where
$A_2=lim_{N,T} (X'MX)^{-1} X'M' \Gamma_1 MX (X'MX)^{-1} NT,$
and
\[\Gamma_1=\begin{bmatrix}
    \Omega_{0}       & \Omega_{1} & \Omega_{2} & \dots & \Omega_{T-1} \\
    \Omega_{1}       & \Omega_{0} & \Omega_{1} & \dots & \Omega_{T-2} \\
    \hdotsfor{5} \\
    \Omega_{T-1}       & \Omega_{T-2} & \Omega_{T-3} & \dots & \Omega_{0}

\vspace{0.5cm}

\end{bmatrix} 
\]
}

\noindent{\bf Remark 9:} The assumption of CLT holds  for $\sum b_t \varepsilon_t$  implies 
$\sum ||\Omega_0^{-1} \Omega_t || < \infty. $ 

\vspace{0.2cm}
\noindent{\bf CASE 2:}  {\it Consider the model in (6) along with assumptions {\bf A1 to A4}.  Further assume that $u_{it}$ 
does not have any cross-sectional dependence but have finite time memory (say of order $q$), or covariance summability property; 
$f_t$ has finite time memory; and   $h_N \to \infty.$ Then
$$ \sqrt\frac{NT}{h_N} (\hat{\beta}_{FE}-\beta) \rightarrow N(0, A_3),$$
where
$A_3=lim_{N,T} (X'MX)^{-1} X'M' \Gamma_2 MX (X'MX)^{-1} \frac{NT} {h_N},$
$\Gamma_2=(\Gamma_{2}(ij))$, 
where $\Gamma_{2}(ij) = \Lambda \Lambda'$, whenever \ $i=j  $,
$\Gamma_{2}(ij) = \Lambda  \theta_k \Lambda'$, whenever \ $0 < |i-j| = k \le q $
and $\Gamma_{2}(ij) = {\bf 0}$ matrix, if \ $|i-j| = k > q$. }
$Var(f_t)= I_r, and  \;\; Cov (f_t, f_{t+k})= \theta_k.$

\vspace{0.2cm}

\noindent{\bf Remark 10:} Covariance summability of $u_t$  includes $Var(u_t)=\Delta_0$, a diagonal matrix. 
Let $Cov (u_t, u_{t+k})= \Delta_k. $ Then Covariance summability implies $\sum|| \Delta_0^{-1} \Delta_k|| <\infty.$  

\vspace{0.2cm}

\noindent{\bf CASE 3:}  {\it  Consider the model in (6) along with assumptions {\bf A1 to A4}.  Further assume that $u_{it}$ does not have any cross-sectional dependence but have finite (say  q) or covariance summable   time memory; $f_t$  is ergodic and has covariance summable with CLT property; and   $h_N \to \infty.$ Then
$$ \sqrt\frac{NT}{h_N} (\hat{\beta}_{FE}-\beta) \rightarrow N(0, A_4),$$
where
$A_4=lim_{N,T} (X'MX)^{-1} X'M' \Gamma_3 MX (X'MX)^{-1} \frac{NT} {h_N},$
\[\Gamma_3=\begin{bmatrix}
    \Lambda \Lambda'  & \Lambda  \theta_1 \Lambda' & \Lambda  \theta_2 \Lambda' & \dots  & \Lambda  \theta_{T-1} \Lambda'  \\
    \Lambda  \theta_1 \Lambda'       & \Lambda \Lambda'  & \Lambda  \theta_1 \Lambda'  & \dots  & \Lambda  \theta_{T-2} \Lambda' \\
    \hdotsfor{5} \\
    \Lambda  \theta_{T-1} \Lambda'      & \Lambda  \theta_{T-2} \Lambda'  & \Lambda  \theta_{T-3} \Lambda' &\dots   & \Lambda \Lambda'  ;
\end{bmatrix} 
\]

$Var(f_t)= I_r, \;\; Cov (f_t, f_{t+k})= \theta_k.$

}

\vspace{0.5cm}

\subsubsection{Estimation of Variance Under Both Cross-sectional and Time Series Dependence: 
Fixed and Random Effect Models}

Here again, it is to be noted that the matrices $A_0, A_1,\ldots A_4$ are unknown in practice and hence need to 
be estimated. The estimator of such matrices can be found easily  in line of Newey and West (1987) or its various 
versions. Here it is worthwhile to mention that if various forms of time dependence, as depicted in this 
sub-section, is known a priori, kernel weight function may be chosen more appropriately. Moreover, the closer 
look of of the Newey-West type estimator suggests that kernel weights excessively injects biases in the variance 
estimator when dependence is pure cross-sectional; no time dependence is present in the data, even when time series
dependence is of short memory type as given in CASE (1A).  Such penalization certainly costs the performance of 
the test based on such estimator, at least in small sample.  

We propose an estimator of $V_{FE}$  as
\[
\bar{V}_{FE} =  \sum_{t=1}^{T} b_t e_t e_t' b_t'+ \sum_{j=1}^{C(T)}\Bigl([\sum_{t=j+1}^{T} b_t e_t e_{t-j}' b_{t-j}'] + [\sum_{t=j+1}^{T} b_t e_t e_{t-j}' b_{t-j}']^{\prime}\Bigr) K(j,T).
\]

Similarly, we propose an estimator of $V_{ols}$ as
\[
\bar{V}_{ols} =  \sum c_t e_t e_t' c_t'+ \sum_{j=1}^{C(T)}\Bigl([\sum_{t=j+1}^{T} c_t e_t e_{t-j}' c_{t-j}'] + [\sum_{t=j+1}^{T} c_t e_t e_{t-j}' c_{t-j}']^{\prime}\Bigr) K(j,T).
\]

$K(.,.)$ is an appropriately chosen kernel depending on the nature of time dependence. At this stage, the appropriate choice of $K(.,.)$ is an empirical issue. Similarly, $C(.)$ is an appropriately chosen truncation lag. For example, $C(.)$ will be zero for strong or semi-strong cross-sectional dependence with finite time memory as in CASE 1A; and $C(.)$ will be $q$ for weak cross-sectional dependence with $MA(q)$ time memory.

\vspace{0.2cm}

\noindent{\bf Remark 11:} Results of Theorem 3, 4, 5 of the previous section 3.1.1 will also hold with respect to the above variance estimators of $\hat{\beta}_{FE}$ and $\hat{\beta}_{ols}$\footnote{For brevity, results are not presented here.}.

\section{Conclusions} Many panel data sets encountered in  economics and social sciences are characterized by cross-sectional dependence. Spatial correlation based on a known weight matrix and error factor structure have been quite familiar in the literature to characterize such dependence. In this paper we formally have defined various forms of error cross-sectional dependence in panel data regression. Several examples have been  given to comprehend the broad scope of such definitions. Results have been  given to show how different forms of dependence are interconnected.   Asymptotic results have been presented (i) for pure cross-sectional dependence; and (ii) also for cross-sectional and time series dependence. Robust standard errors have been provided for popular parameter estimators.

Future research may commence in several directions. First, common shocks are allowed to affect dependent variable only. It may be useful to study the effect of common shocks influencing regressors as well. The models considered by Pesaran (2006) and Bai(2009) may be of interest in this context.  Second, dynamic model with GMM framework may be very useful.    Third, revisiting the estimators in GLS framework would be important. Fourth, it may  be very useful to find tests for various kinds of dependence. Work is in progress in this direction.

\section*{References}

\begin{description}

\item Anderson, B., M. Deistler, A. Filler and C. Zinner (2010):  Generalized linear dynamic factor models—an
approach via singular autoregressions. {\it European Journal of Control}, 3, 211-224

\item Anselin. L. (2001): Spatial Econometrics, {\it in A Companion to Theoretical Econometrics}, ed. by
B. Baltagi. Oxford: Blackwell, 310-330.

\item Bai, J. (2009): Panel Data  Models with  Interactive  Fixed Effects, {\it Econometrica}, 77, 1229-1279.

\item Basak, G. and S. Das (2017): Intercept Homogeneity Test for Fixed Effect Models under Cross-sectional Dependence: Some Insights, {\it Journal of Econometric Methods}, 6, on line version.

\item Baltagi, B. H. and A. Pirotte  (2010): Panel data inference under spatial dependence, {\it Economic Modelling}, 27, 1368-1381. 

\item Banerjee A, Marcellino M, Osbat C (2005): Testing for PPP: should we use panel methods? {\it Empirical Economics}, 30, 77-91.

\item Chudik, A., Pesaran, H., and E. Tosetti (2011): Weak and strong cross
section dependence and estimation of large panels, {\it The Econometrics Journal, },  14,  45-90.

\item  Driscoll., J. C.  and A. C. Kraay (1998): Consistent Covariance Matrix Estimation with
Spatially Dependent Panel Data, {\it Review of Economics and Statistics}, 80, 549-560.

\item Forni, M. and M. Lippi (2001). The generalized factor model: representation theory {\it  Econometric Theory},
17, 1113-1141.

\item Gon\c{c}alves, S. ( 2011): The moving blocks bootstrap for panel linear regression models
with individual fixed-effects. {\it Econometric Theory},  27, 1048-1082.

\item Im, K. S., M. H. Pesaran and S. Shin (2003): Testing for Unit Roots in Heterogeneous Panels, {\it Journal of Econometrics}, 115, 1,  53-74

\item Levin, A., Lin, C., and C.J. Chu (2002): Unit Root Tests in Panel Data: Asymptotic and Finite-sample Properties, {\it Journal of Econometrics}, 108, 1-24.

\item Lewis. W. D (1991): Matrix Theory, World Scientific.

\item Newey, W. K, and K. D. West  (1987).  A Simple, Positive Semi-definite, Heteroskedasticity and Autocorrelation Consistent Covariance Matrix. {\it Econometrica}, 55, 703-708.

\item O$^\prime$ Connell, P. (1998): The over valuation of purchasing power parity, {\it Journal of International
Economics},  44, 1-19.

\item Pesaran H. (2006): Estimation and Inference in Large Heterogeneous Panels With a Multifactor
Error Structure, {\it Econometrica}, 74, 967-1012.

\item Stephan, F.F. (1934): Sampling Errors and Interpretations of Social Data Ordered in
Time and Space.  {\it Journal of the American Statistical Association}, 29, 165-166.

\item Vogelsang,J.T. (2012): Heteroskedasticity, autocorrelation, and spatial correlation robust inference in
linear panel models with fixed-effects, {\it Journal of Econometrics}, 166, 303-315.

\end{description}

\vspace{1cm}

\pagebreak

\noindent {\bf PROOF OF THEOREM 1:} 

Here we will use maximum eigenvalue norm.\\

\noindent RESULT 1: Consider the fixed-effect model. $ \hat{\beta}_{FE} $  is consistent. \\

\noindent{\sc Proof:}

\noindent{\sc Case A: Only with cross-sectional dependence}

$$\begin{array}{lll}
\hat{\beta}_{FE} &= & (X'MX)^{1} X'MY\\
&= &(X'MX)^{-1} X'M(\bar{D}\theta + X\beta + \epsilon)\\
&= & \beta + (X'MX)^{-1} X'M\bar{D}\theta + (X'MX)^{-1} X'M\epsilon. \\
\end{array}
$$
from Assumption 1, we claim that $plim(X'MX)/NT = R$, i.e., we claim that the second order
moment of $MX$ exists.
Now note that,  $||X'MX|| =||X'M M'X|| \le ||M|| \;\; ||X'X||= ||X'X||$.
Hence the matrix R is a finite matrix.

Note that, $(X'MX)^{-1} X'M\bar{D} = 0, \mbox{ where } \bar{D}=[1 1 1 1 1 \ldots  1]'= 1_{NT}.$
This is because $\bar{D}$ is a linear combination of the columns of $D$ and hence
belongs to the column space of $D$ and we know that $MD=0$.

Hence, 
$$\begin{array}{lll}
\hat{\beta}_{FE} &=& \beta +(X'MX)^{-1} X'M \epsilon \\
&=& \beta + [(X'MX)/NT]^{-1} (X'M\epsilon /NT). \\
\end{array}$$
Now we take $plim$ on both sides and we have

$$\begin{array}{lll}
plim \hat{\beta}_{FE} &=& \beta + plim[(X'MX)/NT]^{-1} plim(X'M \epsilon /NT)\\
&=& \beta + [plim(X'MX)/NT]^{-1} plim(X'M \epsilon/NT)\\
&=& \beta + R^{-1} plim(X'M\epsilon /NT). \\
\end{array}$$
Now $limE(X'M\epsilon/NT) = 0$  by Assumption 2.  Again, by Assumption 5,
$$\begin{array}{lll}
||V(X'M \epsilon/NT)|| &=& (1/N^2T^2) ||E [X'M' \; I_T \otimes \Omega \; MX]||\\
&\le & 1/N^2T^2 E||X'M' \; I_T\otimes \Omega \; MX|| \\
&\le & 1/N^2T^2  ||\Omega || \;\; E||X'X||. \\ 
\end{array}$$

Here it may be noted that, for any two $n.n.d.$ matrices $A$ and $B$; $||A \otimes B || \le  ||A || \times ||B||.$ Thus $||I_T \otimes \Omega || \le ||\Omega||.$

Hence for weak dependence, $(V(X'M\epsilon /NT)) \rightarrow 0$  as   $T$  and / or $N \rightarrow \infty $. For intermediate dependence, as $\frac{h_N}{N} \to 0$,  $(V(X'M\epsilon /NT)) \rightarrow 0$ for both fixed  $T$ and for  $T \rightarrow \infty .$
$(V(X'M\epsilon /NT)) \rightarrow 0$ as $T \rightarrow \infty$,  under  strong dependence. 

Hence, $ plim(X'M\epsilon /NT) = 0.$ So we have $ plim \hat{\beta}_{FE} = \beta $ for fixed effect model .

\vspace{2cm}

\noindent{\sc Case B: with both cross-sectional  and time dependence}

Let $\Omega_{t_1, t_2}=\Omega_{t_1- t_2} =Cov (\epsilon_{t_1}, \epsilon_{t_2}).$

Assume that $$\sum_{t=1}^{\infty} ||\Omega_{0}^{-1} \Omega_{t}|| < \infty. $$

Hence, 

\begin{eqnarray}
Var(\epsilon) &=& 
\begin{bmatrix}
    \Omega_{0}       & \Omega_{1} & \Omega_{2} & \dots & \Omega_{T-1} \\
    \Omega_{1}       & \Omega_{0} & \Omega_{1} & \dots & \Omega_{T-2} \\
    \hdotsfor{5} \\
    \Omega_{T-1}       & \Omega_{T-2} & \Omega_{T-3} & \dots & \Omega_{0}
\end{bmatrix} 
 = BAB, \ \mbox{  (say)},
\end{eqnarray}
where 
\[B=
\begin{bmatrix}
    \Omega_{0}^{\frac{1}{2}}       & 0 & 0 & \dots & 0 \\
    0       & \Omega_{0}^{\frac{1}{2}} & 0 & \dots & 0 \\
    \hdotsfor{5} \\
    0       & 0 & 0 & \dots & \Omega_{0}^{\frac{1}{2}}
\end{bmatrix} 
\]
and
\[ A = 
\begin{bmatrix}
    I       & \Omega_{0}^{-\frac{1}{2}} \Omega_{1} \Omega_{0}^{-\frac{1}{2}}   & \Omega_{0}^{-\frac{1}{2}} \Omega_{2} \Omega_{0}^{-\frac{1}{2}} & \dots & \Omega_{0}^{-\frac{1}{2}} \Omega_{T-1} \Omega_{0}^{-\frac{1}{2}} \\
    \Omega_{0}^{-\frac{1}{2}} \Omega_{1} \Omega_{0}^{-\frac{1}{2}}      & I & \Omega_{0}^{-\frac{1}{2}} \Omega_{1} \Omega_{0}^{-\frac{1}{2}}  & \dots & \Omega_{0}^{-\frac{1}{2}} \Omega_{T-2} \Omega_{0}^{-\frac{1}{2}}  \\
    \hdotsfor{5} \\
    \Omega_{0}^{-\frac{1}{2}} \Omega_{T-1} \Omega_{0}^{-\frac{1}{2}}        & \Omega_{0}^{-\frac{1}{2}} \Omega_{T-2} \Omega_{0}^{-\frac{1}{2}}  & \Omega_{0}^{-\frac{1}{2}} \Omega_{T-3} \Omega_{0}^{-\frac{1}{2}} & \dots & I
\end{bmatrix} 
\]

Note that $A_{ii}=I ,\;\; A_{ij}= \Omega_{0}^{-\frac{1}{2}} \Omega_{|i-j|} \Omega_{0}^{-\frac{1}{2}}, \;\; B_{ii}= \Omega_{0}^{-\frac{1}{2}}, \;\; B_{ij}=0, i \ne j.$

$$\begin{array}{lll}
||Var(\epsilon)|| &=& ||BAB|| \\
&\le & ||BB^{\prime}||\; ||A|| \\
&\le & ||\Omega_{0}||  \; max_{i} \sum_{j=0}^{T-1} ||A_{ij}|| \\
& \le & ||\Omega_{0}||\; [ 1+  2\sum_{j=1}^{T-1}   ||\Omega_{0}^{-\frac{1}{2}} \; \Omega_{j} \; \Omega_{0}^{-\frac{1}{2}}||] \\
& \le & ||\Omega_{0}||\; [ 1+ 2\sum_{j=1}^{T-1}   ||\Omega_{0}^{-1} \Omega_{j}||]. \\
\end{array}$$

Therefore, the consistency rate is determined by the $||\Omega_{0}||.$ Hence the result.

\vspace{1cm}
\noindent RESULT 2: Consider the random-effect model. $ \hat{\beta}_{ols}$ is consistent.

\noindent{\sc Case A: Only with cross-sectional dependence}

\noindent{\sc Proof:}
$$\begin{array}{lll}
\hat{\beta}_{ols} &=& (X'\bar{M}X)^{-1} X'\bar{M}Y \\
&=& (X'\bar{M}X)^{-1} X'\bar{M}(\bar{D}\theta + X\beta + \epsilon)\\
&=& \beta + (X'\bar{M}X)^{-1} X'\bar{M}\bar{D}\theta + (X'\bar{M}X)^{-1} X'\bar{M}\epsilon. \\
\end{array} $$

Now it can be easily seen from the form of $\bar{M}= I_{NT} - \bar{D}(\bar{D}'\bar{D})^{-1}\bar{D}'$ that
$\bar{M}\bar{D} = 0. $

Now,
$ \hat{\beta}_{ols} = \beta + (X'\bar{M}X)^{-1} X'\bar{M}\epsilon. $
It is also easy to see  from Assumption 1 that $W=plim[(X'\bar{M}X)/NT] $ is a finite matrix.

$$\begin{array}{lll}
plim  \hat{\beta}_{ols} &=& \beta + plim(X'\bar{M}X)^{-1} X'\bar{M}\epsilon\\
&= & \beta + plim(X'\bar{M}X/NT)^{-1} (X'\bar{M}\epsilon/NT)\\
&=& \beta + W^{-1} plim(X'\bar{M}\epsilon /NT).\\
\end{array} $$

Now $lim E(X'\bar{M}\epsilon /NT) = 0$,    by Assumption 2.  Again, by Assumption 5,

$$\begin{array}{lll}

||V(X'\bar{M} \epsilon/NT)|| &=& 1/N^2T^2 ||E_X (X'\bar{M}' \Omega \bar{M} X)||\\
& \le & 1/N^2T^2 E_X||X'\bar{M}' \Omega \bar{M}X|| \\
& \le & 1/N^2T^2  ||\Omega || \;\; E_X||X'X||. 
\end{array} $$

Hence exactly similar to above result,  $(V(X'M\epsilon /NT)) \rightarrow 0$ as   $T$  and / or $N \rightarrow \infty, $ for weak dependence. For intermediate dependence, as $\frac{h_N}{N} \to 0$,  $(V(X'M\epsilon /NT)) \rightarrow 0, $ for both fixed  $T$ and for  $T \rightarrow \infty .$
$(V(X'M\epsilon /NT)) \rightarrow 0$ as $T \rightarrow \infty$,  under  strong dependence.

Hence, $ plim(X'\bar{M}\epsilon /NT) = 0.$ So we have $ plim \hat{\beta}_{ols} = \beta $ for random effect model . 

\vspace{1cm}

\noindent{\sc Case B: with both cross-sectional  and time dependence}\\
\vspace{1cm}
As in Result 1.

\vspace{1cm}

\noindent {\bf PROOF OF THEOREM 2:}\\

\vspace{0.4cm}

\noindent {\bf Lemma 1:}
{ Define a matrix  $V_F = E((\varepsilon_t \varepsilon_t') \otimes (\varepsilon_t \varepsilon_t'))$.  Then $trace(V_F) \ge \lambda_{\max}(V_F) \ge  \sum_{i,j}\omega^2_{ij}  \ge
(\lambda_{\max}(\Omega \otimes \Omega)) \ge \lambda_{\max} (\Omega^2) = \lambda_{\max}^2 (\Omega) $. }

\vspace{0.4cm}

\noindent {\bf PROOF OF Lemma 1:}

Note that $V_F$ is  based on fourth moments and cross moments whose dimension is $ N^2 \times N^2. $ 

Note, $A_1 A_2 \otimes B_1 B_2 = (A_1 \otimes B_1)(A_2 \otimes B_2)$
(see C.R.Rao, Linear Statistical Inference and Its Applications, p.29) whenever their orders are compatible for multiplication.
Hence
\begin{eqnarray}
V_F &=& E(\epsilon_t \epsilon_t' \otimes \epsilon_t \epsilon_t') \nn\\
&=& E[(\epsilon_t \otimes \epsilon_t)(\epsilon_t' \otimes  \epsilon_t')] .
\end{eqnarray}
For two symmetric non-negative definite (n.n.d.) matrices $A$ and $B$
with $A \ge B$, i.e., $A - B$ non-negative definite,
one has $\lambda_{\max}(A) \ge \lambda_{\max}(B)$.
Now note,
$$
E\Bigl[[(\epsilon_t \otimes \epsilon_t) - E(\epsilon_t \otimes \epsilon_t)] [(\epsilon_t' \otimes  \epsilon_t') - E(\epsilon_t' \otimes \epsilon_t')]\Bigr]
$$
is an n.n.d. matrix and it is equal to 
$$
E[(\epsilon_t \otimes \epsilon_t)(\epsilon_t' \otimes  \epsilon_t')] - [E(\epsilon_t \otimes \epsilon_t)][E(\epsilon_t' \otimes \epsilon_t')] .
$$
Thus,
\begin{eqnarray}
\lambda_{\max}(V_F) &\ge & \lambda_{\max}([E(\epsilon_t \otimes \epsilon_t)][E(\epsilon_t' \otimes \epsilon_t')]) \nn\\
&=& [E(\epsilon_t' \otimes \epsilon_t')][E(\epsilon_t \otimes \epsilon_t)] \nn\\
&=& \sum_{i,j} \omega_{ij}^2 \nn\\
&=& Trace(\Omega^2) \nn\\
&\ge & \lambda_{max} (\Omega^2) \nn\\
&=& (\lambda_{\max} (\Omega))^2 .
\end{eqnarray}

Hence for strong dependence $\lambda_{\max}(V_F) = O(N^2). $ This is because $\lambda_{\max}(V_F) \le  O(N^2)$, but for strong dependence, $\lambda_{max} (\Omega)=O(N).$  For weak dependence, in general, $O(\lambda_{\max}(V_F)) \ge O(N)$. This is because $Trace (\Omega^2) \ge O(N).$  \qed

\vspace{0.4cm}
{\bf Proof of Theorem 2 (contd.):}

Treating $X$  as constants, we  apply Liapounav CLT on the sequence of independent random variables $\{ W_t= l^{\prime}b_t \varepsilon_t : 1 \le t\le T\}$. Thus we prove, 
$$\sum E(|W_t|^{2+\delta})/(\sum{Var(W_t}))^{1+\delta/2} \to 0  \;\; \mbox{as} \;\; T \to \infty. $$
For $\delta=2$, consider the quantity, $\sum E_X(l' b_t \varepsilon_t \varepsilon_t' b_t' l) ^2$,  and $ l'l=1. $  This expectation is conditional on $X$.

Thus,
$$\sum E_X(l' b_t \varepsilon_t \varepsilon_t' b_t' l) ^2  \le \sum (l' b_t  b_t' l) ^2 \lambda_{\max}(V_F) 
$$
where
$$\begin{array}{lll}
\sum (l' b_t  b_t' l) ^2  & = & \sum O(\frac{1} {NT^2})^2 ,\\
&=& O (\frac{1}{N^2 T^{3}}). \end{array} $$ 

For strong dependence, if the eigenvector  corresponding to the largest eigenvalue of $\Omega$ belongs to the row-space of 
 $(\sum b_t' b_t)$ then

$$\begin{array}{lll}
 (\sum l' b_t \Omega b_t' l) ^2  &= & \{(\sum l' b_t b_t' l) O(\lambda_{\max} (\Omega))\}^2,\\
 &=& O (\frac{\lambda_{\max} (\Omega)} {NT})^2 ,  
\end{array} $$ 
and hence, 
$$\begin{array}{lll}
\frac{\sum E_X(l' b_t \varepsilon_t \varepsilon_t' b_t' l) ^2 }{ (\sum l' b_t \Omega b_t' l) ^2} 
  &=& O (\frac{1} {T}).  
\end{array} $$ 

For weak dependence, 
$$\begin{array}{lll}
 (\sum l' b_t \Omega b_t' l) ^2  &\ge & \{(\sum l' b_tb_t' l) (\lambda_{\min} (\Omega)\}^2,\\
 &=& O (\frac{1} {NT})^2 ,  
\end{array} $$ 
since $\lambda_{\min} (\Omega)=O(1).$

Therefore,
$$\begin{array}{lll}
\frac{\sum E_X(l' b_t \varepsilon_t \varepsilon_t' b_t' l) ^2 }{ (\sum l' b_t \Omega b_t' l) ^2} 
  &=& O (\frac{N} {T}) ,  
\end{array} $$ 
as $\lambda_{\max}(V_F) = O(N)$.

\vspace{0.5cm}

\noindent {\bf PROOF OF THEOREM 4:}

Proof is exactly similar to that of \noindent {\bf Theorem 2}.

\vspace{0.5cm}

\noindent {\bf PROOF OF THEOREM 3 and 5:}

Proof of Theorem 3 and 5  is avoided for brevity. Proof of Theorem 3 is similar to that of Basak and Das (2017).

\end{document}